\documentclass{moriond}

\usepackage{amssymb}
\usepackage{lineno}

\def\Journal#1#2#3#4{{#1} {\bf #2}, #3 (#4)}

\def\NIMA{{\em Nucl. Instrum. Methods} A}

\def\PLB{{\em Phys. Lett.}  B}

\def\EPJC{{\em Eur. Phys. J.} C}
\def\CHPC{{\em Chin. Phys.} C}

\def\be{\begin{equation}}
\def\ee{\end{equation}}
\def\bea{\begin{eqnarray}}
\def\eea{\end{eqnarray}}

\newcommand{\Photo}{}

\newcommand{\kpimmws}{K^{\pm} \to \pi^{\mp} \mu^{\pm} \mu^{\pm}}
\newcommand{\kpimmns}[1]{K^{#1} \to \pi \mu \mu}
\newcommand{\kpimm}[1]{K^{#1} \to \pi^{#1} \mu^{+} \mu^{-}}

\newcommand{\kmutwoN}[1]{K^{#1} \to \mu^{#1} N_{4}}
\newcommand{\kpichi}[1]{K^{#1} \to \pi^{#1} \chi}
\newcommand{\Npimuws}{N_{4} \to \pi^{\mp}\mu^{\pm}}
\newcommand{\Npimurs}{N_{4} \to \pi^{\pm}\mu^{\mp}}
\newcommand{\chimumu}{\chi \to \mu^{+}\mu^{-}}

\newcommand{\kthreepic}[1]{K^{#1} \to \pi^{#1} \pi^+ \pi^-}
\newcommand{\kthreepin}[1]{K^{#1} \to \pi^{#1} \pi^0 \pi^0}
\newcommand{\kpmm}{K_{\pi\mu\mu}}
\newcommand{\kpmmlnv}{K_{\pi\mu\mu}^{\rm LNV}}
\newcommand{\kpmmlnc}{K_{\pi\mu\mu}^{\rm LNC}}
\newcommand{\npmmlnv}{N_{\pi\mu\mu}^{\rm LNV}}
\newcommand{\kmufour}[1]{K^{#1} \to \pi^{+} \pi^{-} \mu^{#1} \nu}
\newcommand{\kmunumumu}[1]{K^{#1} \to \mu^{+} \mu^{-} \mu^{#1} \nu}
\newcommand{\kpinn}[1]{K^{#1} \to \pi^{#1} \nu \bar{\nu}}
\newcommand{\eqdef}{\stackrel{\,def}{=}}

\begin{document}
\vspace*{4cm}
\title{SEARCHES FOR LEPTON NUMBER VIOLATION AND RESONANCES IN THE $K^{\pm}\rightarrow \pi \mu \mu$ DECAYS AT THE NA48/2 EXPERIMENT}

\author{K. MASSRI$\phantom{^{\,}}$\footnote{On behalf of the NA48/2 Collaboration: G.~Anzivino, R.~Arcidiacono, W.~Baldini, S.~Balev, J.R.~Batley, M.~Behler, S.~Bifani, C.~Biino, A.~Bizzeti
% ...
}}

\address{Department of Physics, University of Liverpool, The Oliver Lodge Laboratory,\\Liverpool L69 7ZE,  United Kingdom}

\maketitle\abstracts{The NA48/2 experiment at CERN collected a large sample of charged kaon decays into final states with multiple charged particles in 2003--2004. A new upper limit on the rate of the lepton number violating decay $\kpimmws$ obtained from this sample is reported: $\mathcal{B}(\kpimmws)<8.6 \times 10^{-11}$ at 90\% CL. Searches for two-body resonances in the $\kpimmns{\pm}$ decays (including heavy neutral leptons~$N_4$ and inflatons~$\chi$) in the accessible range of masses and lifetimes are also presented.
In the absence of a signal, upper limits are set on the products of branching ratios~$\mathcal{B}(\kmutwoN{\pm})\mathcal{B}(\Npimuws)$ and $\mathcal{B}(\kpichi{\pm})\mathcal{B}(\chimumu)$ as functions of the resonance mass and lifetime. These limits are in the $10^{-10}-10^{-9}$ range for resonance lifetimes below 100~ps.}

\section{Introduction}
Neutrinos are strictly massless within the Standard Model~(SM), due to the absence of right-handed neutrino states. However, since the observation of neutrino oscillations~\cite{pdg} has unambiguously demonstrated the massive nature of neutrinos, right-handed neutrino states must be included.
A natural extension of the SM involves the inclusion of sterile neutrinos which mix with ordinary neutrinos to explain several open questions. An example of such a theory is the Neutrino Minimal Standard Model~($\nu$MSM)~\cite{as05_02}.
In this model, three massive right-handed neutrinos are introduced to explain simultaneously neutrino oscillations, dark matter and baryon asymmetry of the Universe: the lightest has mass~$\mathcal{O}(1\mbox{ keV})$ and is a dark matter candidate; the other two, with masses ranging from 100~MeV/$c^2$ to few GeV/$c^2$, are responsible for the masses of the SM neutrinos (via see-saw mechanism) and introduce extra CP violating phases to account for baryon asymmetry.
The $\nu$MSM can be further extended by adding a real scalar field, to incorporate inflation and provide a common source for electroweak symmetry breaking and for right-handed neutrino masses~\cite{sh06}.
These SM extensions predict new particles, such as heavy neutrinos and inflatons, which could be produced in $\kpimmns{\pm}$ decays.
In particular, the Lepton Number Violating~(LNV) $\kpimmws$ decay, which is forbidden in the SM, could proceed via the production of on-shell Majorana neutrinos~\cite{at09}, while inflatons~$\chi$ could be produced in $K^{\pm}\to\pi^{\pm}\chi$ decays, and promptly decaying to $\chi\to\mu^+\mu^-$~\cite{be10,be14}.

The large statistics of the samples of charged kaon decays into final states with multiple charged particles collected in 2003--2004 by the NA48/2 experiment at CERN allows to search for the forbidden LNV $\kpimmws$ decay, as well as for two-body resonances in $\kpimmns{\pm}$ decays.
Since a particle~$X$ produced in a $K^{\pm}\to\mu^{\pm}X$ ($K^{\pm}\to\pi^{\pm}X$) decay and decaying promptly to $\pi^{\pm}\mu^{\mp}$ ($\mu^+\mu^-$) would produce a narrow spike in the invariant mass $M_{\pi\mu}$ ($M_{\mu\mu}$) spectrum, the invariant mass distributions of the collected $\kpimmns{\pm}$ samples have been scanned looking for such a signature.
The results of these searches are presented, together with the prospects for the search for Lepton Number and Flavour Violation~(LNFV) at the NA62 experiment, which aims to precisely measure the branching ratio of the $K^{+}\to\pi^+\nu\bar{\nu}$ decay.

\section{Experimental Apparatus and Data Taking conditions}
\label{sec:detector}
The NA48/2 experiment at CERN SPS was a multi-purpose $K^{\pm}$ experiment which collected data in 2003--2004, whose main goal was to search for direct CP violation in the $\kthreepic{\pm}$ and $\kthreepin{\pm}$ decays~\cite{ba07}.
Simultaneous and collinear $K^+$ and $K^-$ beams of the same momentum ($60\pm3.7$)~GeV/$c$ were produced by the 400~GeV/$c$ SPS primary proton beam, which impinged on a Beryllium target, and were steered into a 114~m long decay region, contained in a vacuum (at pressure $< 10^{-4}$~mbar) cylindrical tank.
The downstream part of the vacuum tank was sealed by a convex Kevlar window, that separated the vacuum from the helium at atmospheric pressure in which a magnetic spectrometer, formed of 4 drift chambers (DCHs) and a dipole magnet providing a horizontal momentum kick~$p_t = 120$~MeV$/c$, was installed.
The spatial resolution of each DCH was $\sigma_x = \sigma_y = 90$~$\mu$m, while the momentum resolution of the spectrometer was~$\sigma(p)/p = (1.02 \oplus 0.044 \cdot p)\%$, where the momentum~$p$ is measured in GeV/$c$.
A hodoscope~(HOD)
was placed downstream of the spectrometer and provided fast signals for trigger purposes, as well as time measurements for charged particles with a resolution of $\sim 300$~ps.
The HOD was followed by a LKr electromagnetic calorimeter
with a depth of 127~cm, corresponding to 27 radiation lengths.
The front plane had an octagonal shape and was segmented in 13248 cells with size $2 \times 2$~cm$^2$.
The LKr calorimeter energy resolution was measured to be~$\sigma_E/E = (3.2/\sqrt{E} \oplus 9.0/E \oplus 0.42)\%$,
where $E$ is the energy expressed in GeV. The space resolution~$\sigma_{x,y}$ of the LKr was~$\sigma_{x,y} = (4.2/\sqrt{E} \oplus 0.6)~\mbox{mm}$, and the time resolution on the single shower was $\sigma_t = 2.5\mbox{ ns}/\sqrt{E}$.
The LKr was followed by a hadronic calorimeter (not used for the present measurement) and a muon detector~(MUV). The MUV consisted of three $2.7\times2.7$~m$^2$ planes of plastic scintillator strips, each preceded by a 80~cm thick iron wall and alternately aligned horizontally and vertically.
The strips were 2.7~m long and 2~cm thick, and they were read out by photomultipliers at both ends. The widths of the strips were 25~cm in the first two planes, and 45~cm in the third plane. A detailed description of the NA48/2 beam line and the detector layout can be found in Refs.~\cite{ba07,fa07}.



\section{Event reconstruction and selection}
\label{sec:selection}
The event selection is based on the reconstruction of a three-track vertex: given the resolution of the vertex longitudinal position ($\sigma_{vtx} = 50$~cm), $\kpimmws$ and $\kpimm{\pm}$ decays (denoted $\kpmmlnv$ and $\kpmmlnc$ below) mediated by a short-lived ($\tau\lesssim 10$~ps) resonant particle are indistinguishable from a genuine three-track decay.
The size of the selected $\kpmm$ samples is normalised relative to the abundant $K^\pm\to\pi^\pm\pi^+\pi^-$ channel (denoted $K_{3\pi}$ below).
The $\kpmm$ and $K_{3\pi}$ samples are collected concurrently using the same trigger logic. Since the $\mu^\pm$ and $\pi^\pm$ masses are close ($m_\mu/m_\pi=0.76$), the signal and the normalisation final states have similar topologies. This leads to first order cancellation of the systematic effects induced by imperfect kaon beam description, local detector inefficiencies, and trigger inefficiency.

The selections for the $\kpmm$ and $K_{3\pi}$ modes have a large common part: a vertex satisfying the following principal criteria is required.
\begin{itemize}
\vspace{-0.1cm}
\item[-] The total charge of the three tracks is $Q=\pm1$.
\vspace{-0.2cm}
\item[-] The vertex longitudinal position is within the 98~m long fiducial decay volume (i.e. downstream the final collimator).
\vspace{-0.2cm}
\item[-] The vertex tracks must have momentum~$p$ within the range $(5;55)~{\rm GeV}/c$, the total momentum of the three tracks $|\sum\vec p_i|$ must be consistent with the beam nominal range $(55;65)~{\rm GeV}/c$, and the total transverse momentum of the three tracks with respect to the beam axis direction is $p_T<10^{-2}~{\rm GeV}/c$.
\vspace{-0.1cm}
\end{itemize}
If several vertices fulfill the above conditions, the one with the lowest fit $\chi^2$ is considered.
The vertex tracks are required to be consistent in time and to be in DCH, HOD, LKr and MUV geometric acceptances. Track separations are required to exceed 2~cm in the DCH1 plane to suppress photon conversions, and 20~cm in the LKr, MUV1 and MUV2 front planes to minimize particle misidentification due to shower overlaps and to multiple Coulomb scattering.

The following criteria are used to select the $\kpmmlnv$ ($\kpmmlnc$) candidates.
\begin{itemize}
\vspace{-0.1cm}
\item[-] The vertex must be composed of one $\pi^\pm$ candidate (with the ratio of energy~$E$ in the LKr calorimeter to momentum~$p$ measured in the spectrometer $E/p<0.95$ to suppress electrons, and no in-time associated hits in the MUV), and a pair of identically (oppositely) charged $\mu^\pm$ candidates (with $E/p<0.2$ and associated hits in the first two planes of the MUV). The $\pi^{\pm}$ candidate is required to have momentum above $15~{\rm GeV}/c$ to ensure high muon rejection efficiency. The muon identification efficiency has been measured to be above $99\%$ for $p>15$~GeV/$c$.
\vspace{-0.2cm}
\item[-] The invariant mass of the three tracks in the $\pi^\mp\mu^\pm\mu^\pm$ ($\pi^\pm\mu^+\mu^-$) hypothesis must satisfy the requirement
$|M_{\pi\mu\mu}-M_K|< 5~{\rm MeV}/c^2$ ($|M_{\pi\mu\mu}-M_K|< 8~{\rm MeV}/c^2$), where $M_K$ is the nominal $K^{\pm}$ mass~\cite{pdg}. This range corresponds to $\pm 2$ ($\pm3.2$) times the resolution on $M_{\pi\mu\mu}$.
\vspace{-0.1cm}
\end{itemize}
An additional requirement is applied to the $\kpmm$ samples, when searching for resonances:
\begin{itemize}
\vspace{-0.1cm}
\item[-] $|M_{ij}-M_{X}|<2\sigma(M_{ij})$, where $M_{ij}$ is the invariant mass of the $ij$ pair ($ij = \pi^{\pm}\mu^{\mp}, \mu^+\mu^-$), $M_X$ is the assumed resonance mass, and $\sigma (M_{ij})$ is the resolution on the invariant masses~$M_{ij}$.
\vspace{-0.6cm}
\end{itemize}
Independently, the following criteria are applied to select the $K_{3\pi}$ sample.
\begin{itemize}
\vspace{-0.1cm}
\item[-] The pion identification criterion described above is applied to the odd-sign pion only, to symmetrize the selection of the signal and normalisation modes.
\vspace{-0.2cm}
\item[-] The invariant mass of the three tracks in the $3\pi^\pm$ hypothesis is in the range $|M_{3\pi}-M_K|<5~{\rm MeV}/c^2$. This interval corresponds approximately to $\pm 3$ times the resolution on~$M_{3\pi}$.
\vspace{-0.1cm}
\end{itemize}
No restrictions are applied to additional energy depositions in the LKr calorimeter, nor to extra tracks not belonging to the three-track vertex, to decrease the sensitivity to accidental activity.

\boldmath
\section{Selected data samples}
\begin{figure}[h]
\begin{minipage}{0.5\textwidth}
\includegraphics[width=\textwidth]{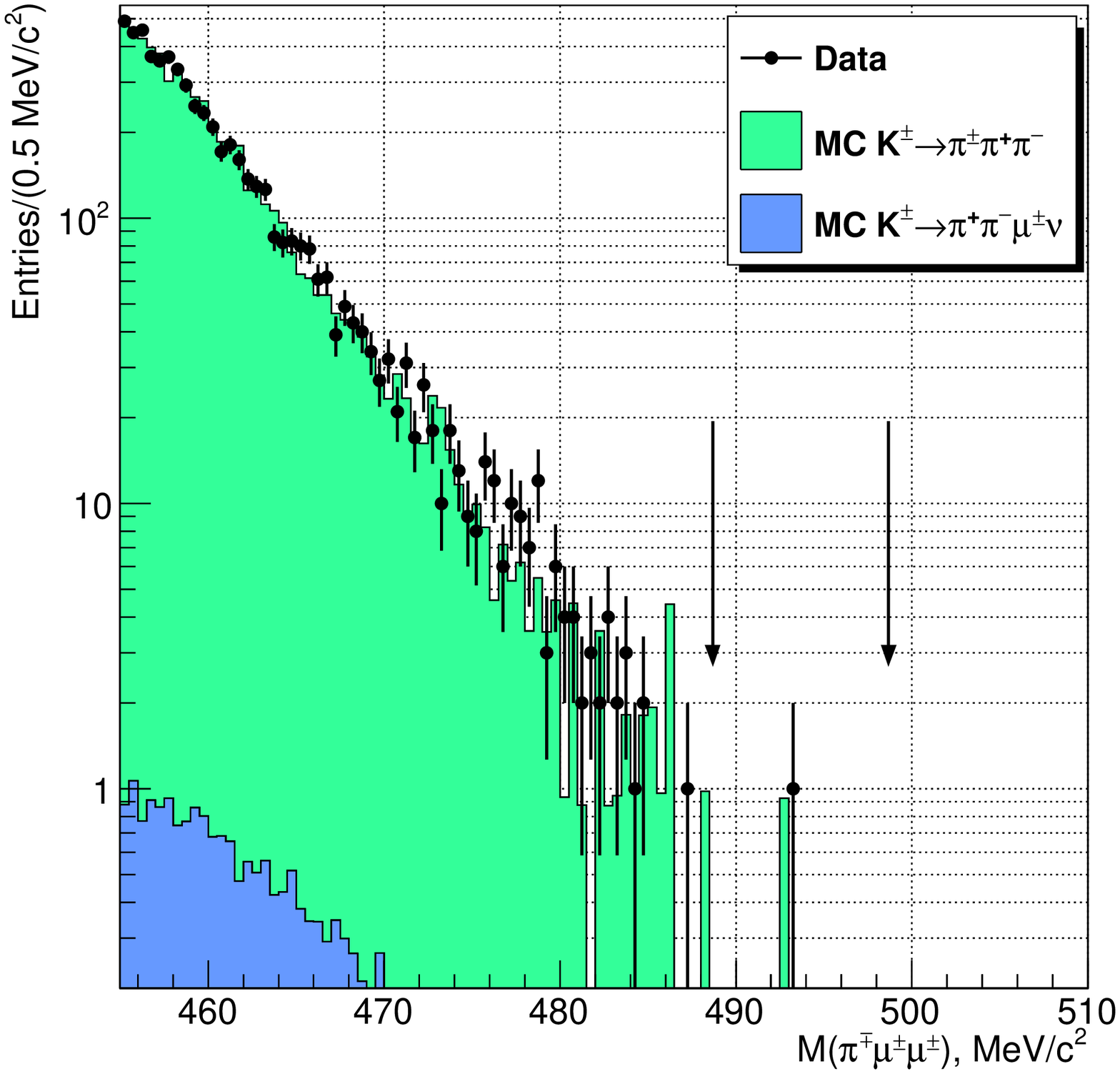}%
\end{minipage}
\hfill
\begin{minipage}{0.5\textwidth}
\includegraphics[width=\textwidth]{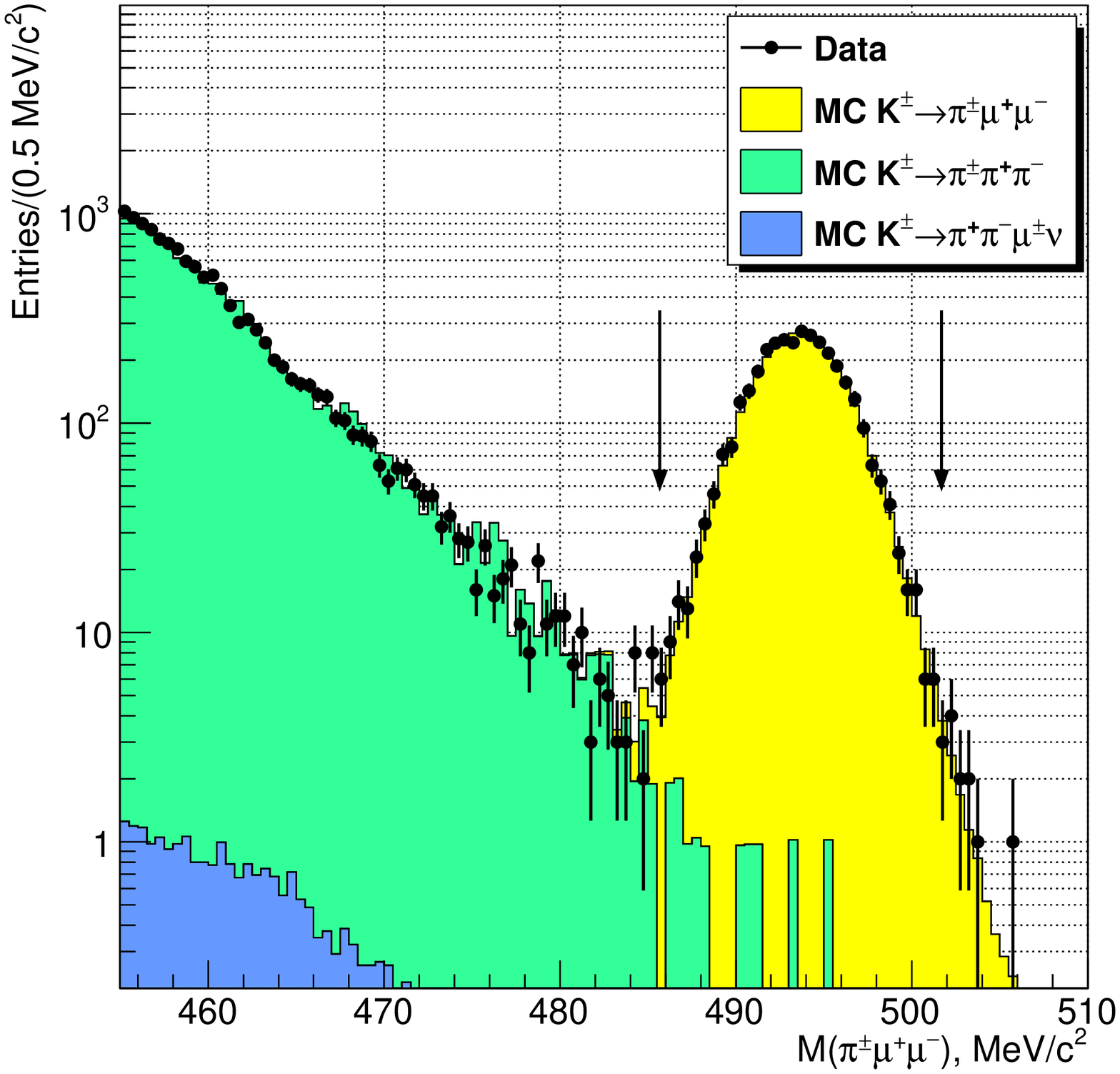}%
\end{minipage}
\caption{Invariant mass distributions of data and MC events passing the $\kpmmlnv$ (left) and
$\kpmmlnc$~(right) selections. The signal mass regions are indicated with vertical arrows.}\label{fig:mpimm}  
\end{figure}

\unboldmath
\label{sec:datasamples}
The invariant mass distributions of data and MC events passing the $\kpmmlnv$ and $\kpmmlnc$ selections are shown in Fig.~\ref{fig:mpimm}. The signal mass regions are indicated with vertical arrows.

\noindent One event is observed in the signal region after applying the $\kpmmlnv$ selection,
while 3489 $\kpmmlnc$ candidates are selected with the $\kpmmlnc$ selection.
The number of expected background events in the $\kpmmlnv$ sample is $N_{bkg} = 1.163 \!\pm\! 0.867_{stat} \!\pm\! 0.021_{ext} \!\pm\! 0.116_{syst}$,
the main source of which is the $K_{3\pi}$ decay with two subsequent $\pi^{\pm}\to\mu^{\pm}\nu$ decays.
The estimated $K_{3\pi}$ background contamination in the $\kpmmlnc$ sample is $(0.36\pm0.10)\%$, based on MC simulations. Such a level of purity allows to consider the $\kpmmlnc$ decay as the only background for the resonance searches over the collected $\kpmmlnc$ sample.

\noindent The number of $K^\pm$ decays in the 98~m long fiducial decay region is computed as
\begin{equation}
N_K = \frac{N_{3\pi}\cdot D}{{\cal B}(K_{3\pi})A(K_{3\pi})} = (1.64\pm0.01)\times 10^{11},
\end{equation}
where $N_{3\pi} = 1.37\times10^{7}$ is the number of data candidates reconstructed within the $K_{3\pi}$ selection, $D = 100$ is the downscaling factor of the considered $K_{3\pi}$ sample, ${\cal B}(K_{3\pi})$ is the nominal branching ratio of the $K_{3\pi}$ decay mode~\cite{pdg} and $A(K_{3\pi}) = (14.955 \pm 0.004)\%$ is the acceptance of the $K_{3\pi}$ selection for the $K_{3\pi}$ decays evaluated with MC simulations.
\boldmath
\section{Search for two-body resonances}
\begin{figure}[h]
\begin{minipage}{0.333\textwidth}
\includegraphics[width=\textwidth]{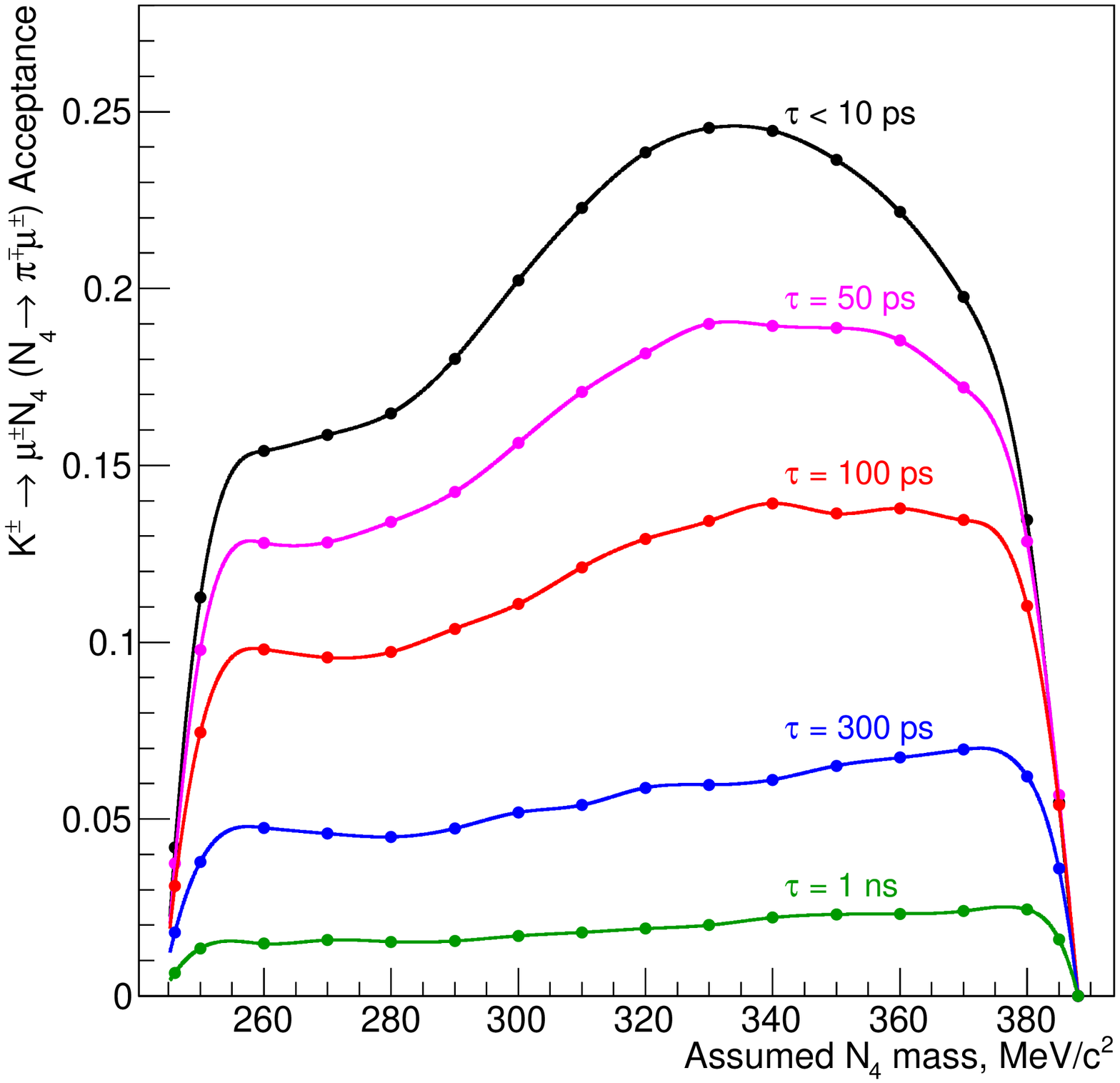}%
\end{minipage}
\put(-14,62){\Large\bf a}
\hfill
\begin{minipage}{0.333\textwidth}
\includegraphics[width=\textwidth]{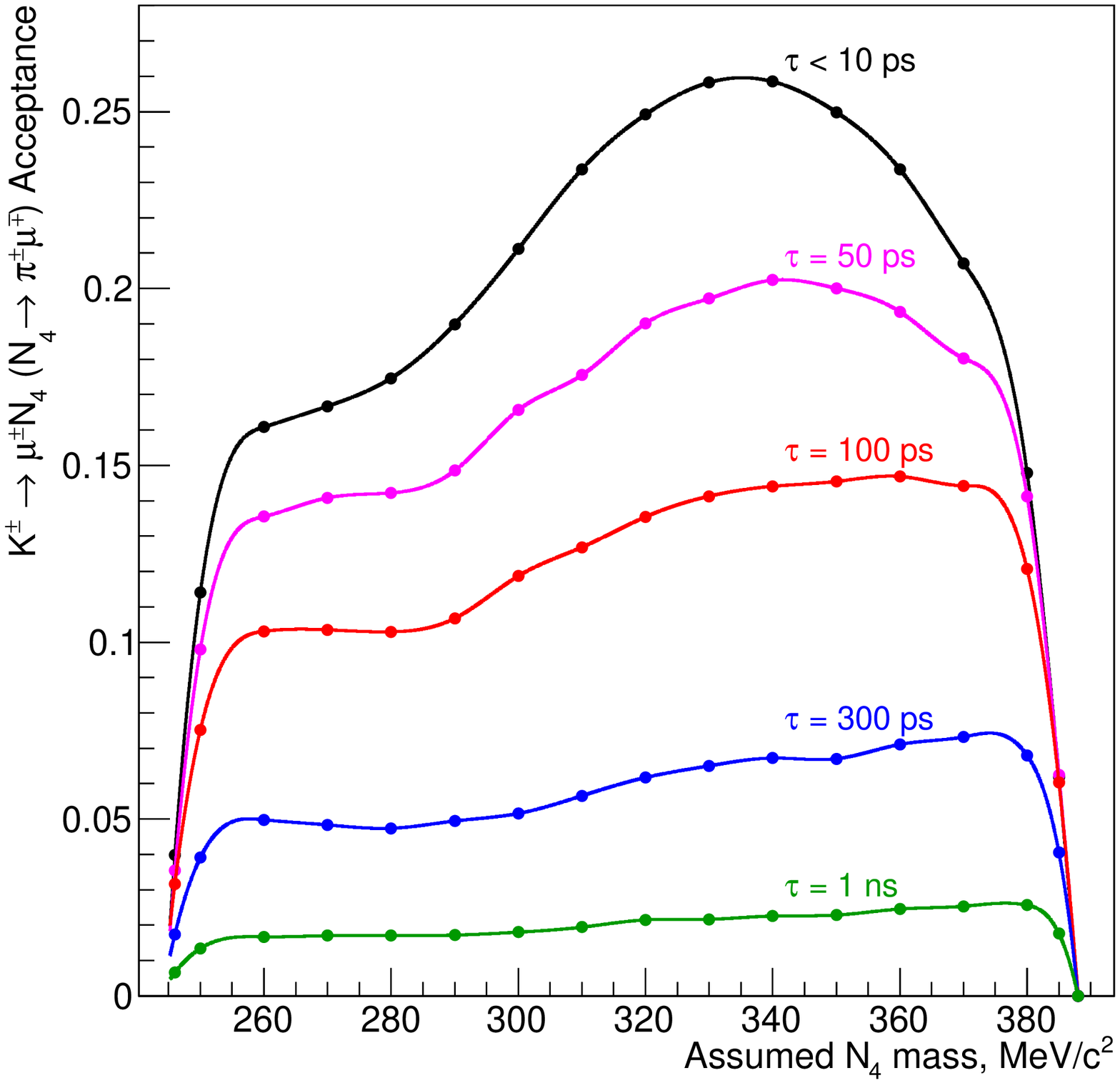}%
\end{minipage}
\put(-14,62){\Large\bf b}
\begin{minipage}{0.333\textwidth}
\includegraphics[width=\textwidth]{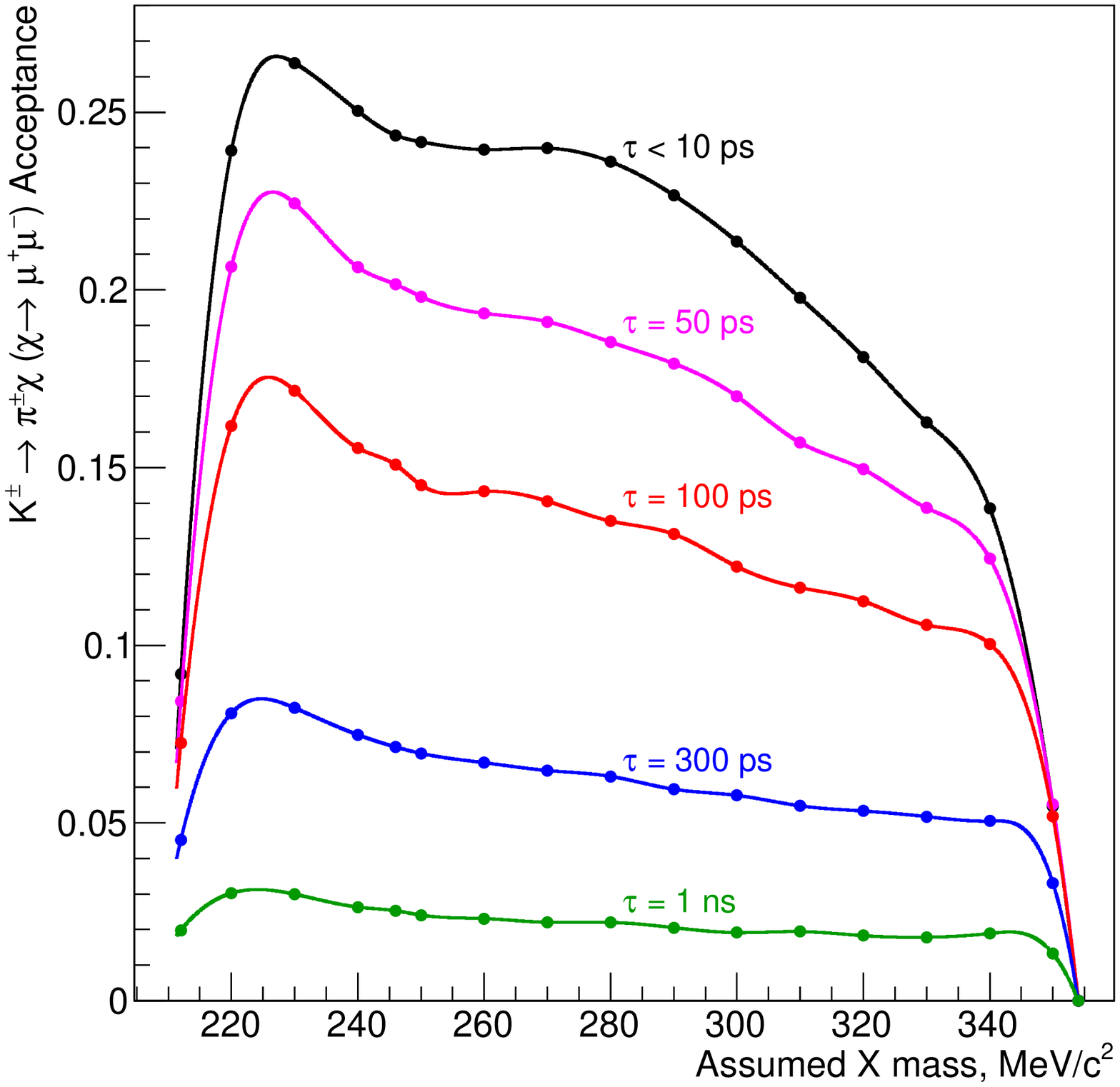}%
\end{minipage}
\put(-14,62){\Large\bf c}
\caption{Acceptances as a function of the resonance mass and lifetime of a) the $\kpmmlnv$ selection for $K^{\pm}\to\mu^{\pm}N_4$ decays followed by the $N_4\to\pi^{\mp}\mu^{\pm}$ decay; b) the $\kpmmlnc$ selection for $K^{\pm}\to\mu^{\pm}N_4$ decays followed by the $N_4\to\pi^{\pm}\mu^{\mp}$ decay; c) the $\kpmmlnc$ selection for $K^{\pm}\to\pi^{\pm}X$ decays followed by the $X\to\mu^+\mu^-$ decay.
Due to the required three-track topology of the selected events, the acceptances scale as $1/\tau$ for resonance lifetimes $\tau>1$~ns.}\label{fig:acc}  
\end{figure}

\unboldmath
\label{sec:res_search}
A peak search assuming different mass hypotheses is performed over the distributions of the invariant masses $M_{ij}$ ($ij = \pi^{\pm}\mu^{\mp}, \mu^+\mu^-$) of the selected $\kpmm$ samples. The precise evaluation of the acceptance for $K^{\pm}\to\mu^{\pm}X$ ($K^{\pm}\to\pi^{\pm}X$) decays with subsequent $X\to\pi^{\pm}\mu^{\mp}$ ($X\to\mu^+\mu^-$) 
decay as a function of the resonance mass and lifetime has been performed with dedicated MC simulations and it is shown in Fig.~\ref{fig:acc}.

\noindent The mass steps of the resonance searches and the width of the signal mass windows around the assumed mass are determined by the resolutions $\sigma(M_{ij})$ on the invariant masses $M_{ij}$ ($ij = \pi^{\pm}\mu^{\mp}, \mu^+\mu^-$): the mass step is set to be equal to $\sigma(M_{ij})/2$, while the half-width of the signal mass window is $2 \sigma(M_{ij})$. 
Therefore, the results obtained in the neighbouring mass hypotheses are highly correlated, as the signal mass window is about 8 times wider than the mass step of the resonance scan.
In total, 284 (267) and 280 mass hypotheses are tested respectively for the search of resonances in the $M_{\pi\mu}$ distribution of the $\kpmmlnv$ ($\kpmmlnc$) candidates and in the $M_{\mu\mu}$ distribution of the $\kpmmlnc$ candidates, covering the full kinematic ranges.

\begin{figure}[p]
\begin{center}
\begin{minipage}{0.5\textwidth}
\includegraphics[width=\textwidth]{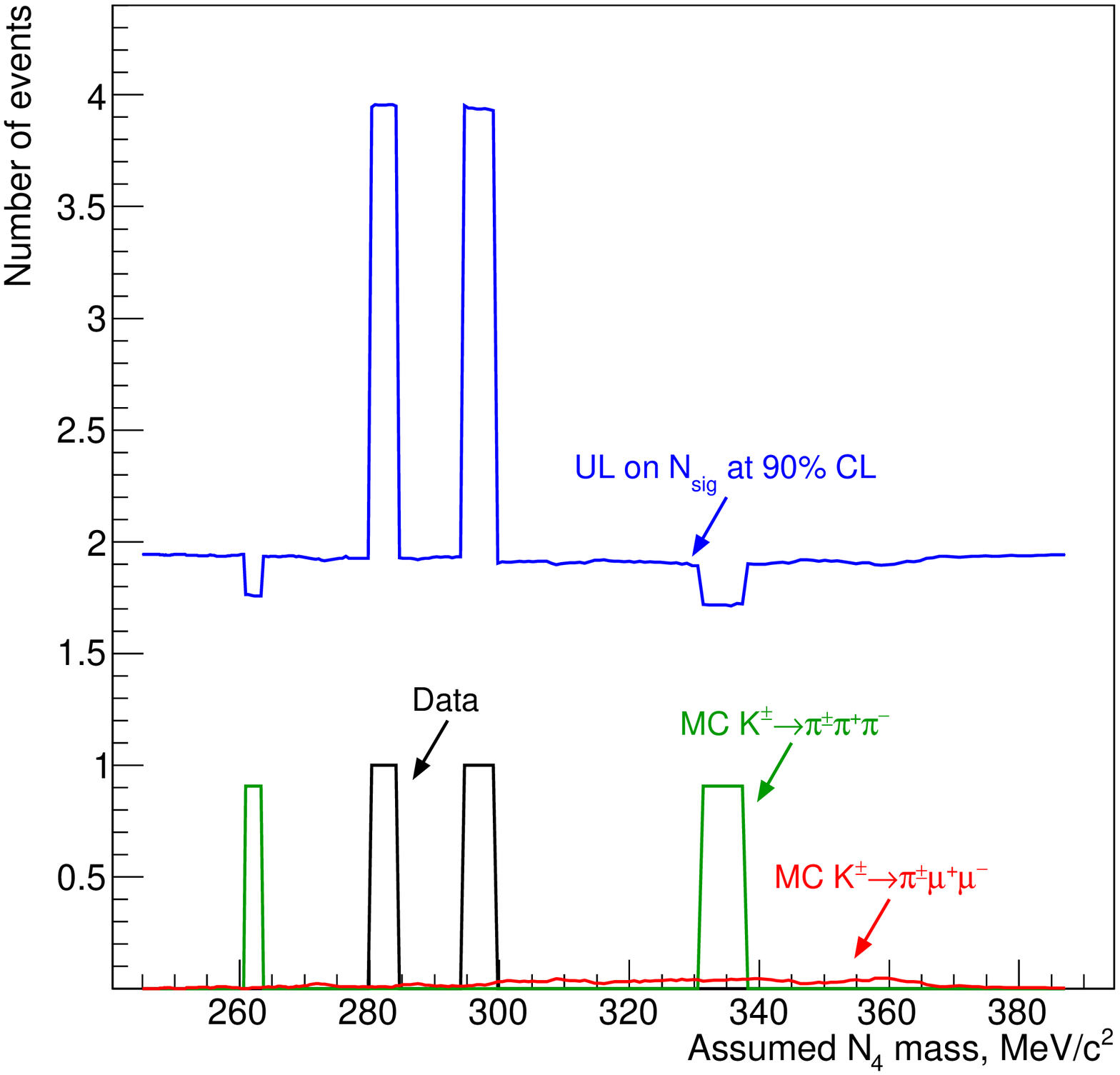}\\%
\hspace*{0.02\textwidth}\includegraphics[width=0.98\textwidth]{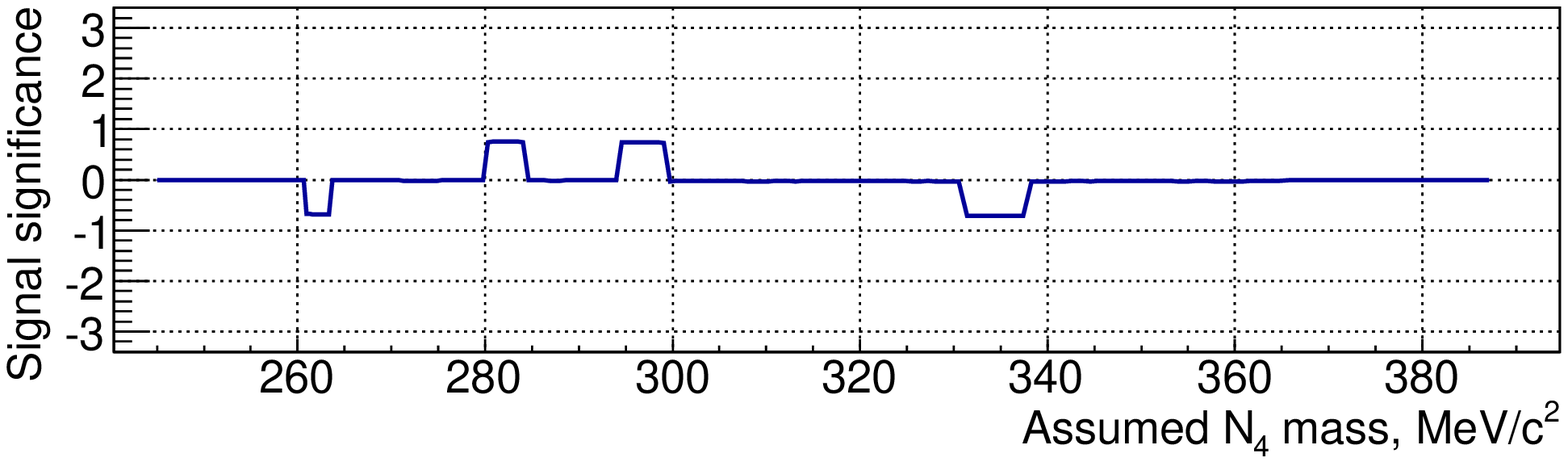}%
\end{minipage}
\put(-14,128){\Large\bf a}
\hfill
\begin{minipage}{0.5\textwidth}
\includegraphics[width=\textwidth]{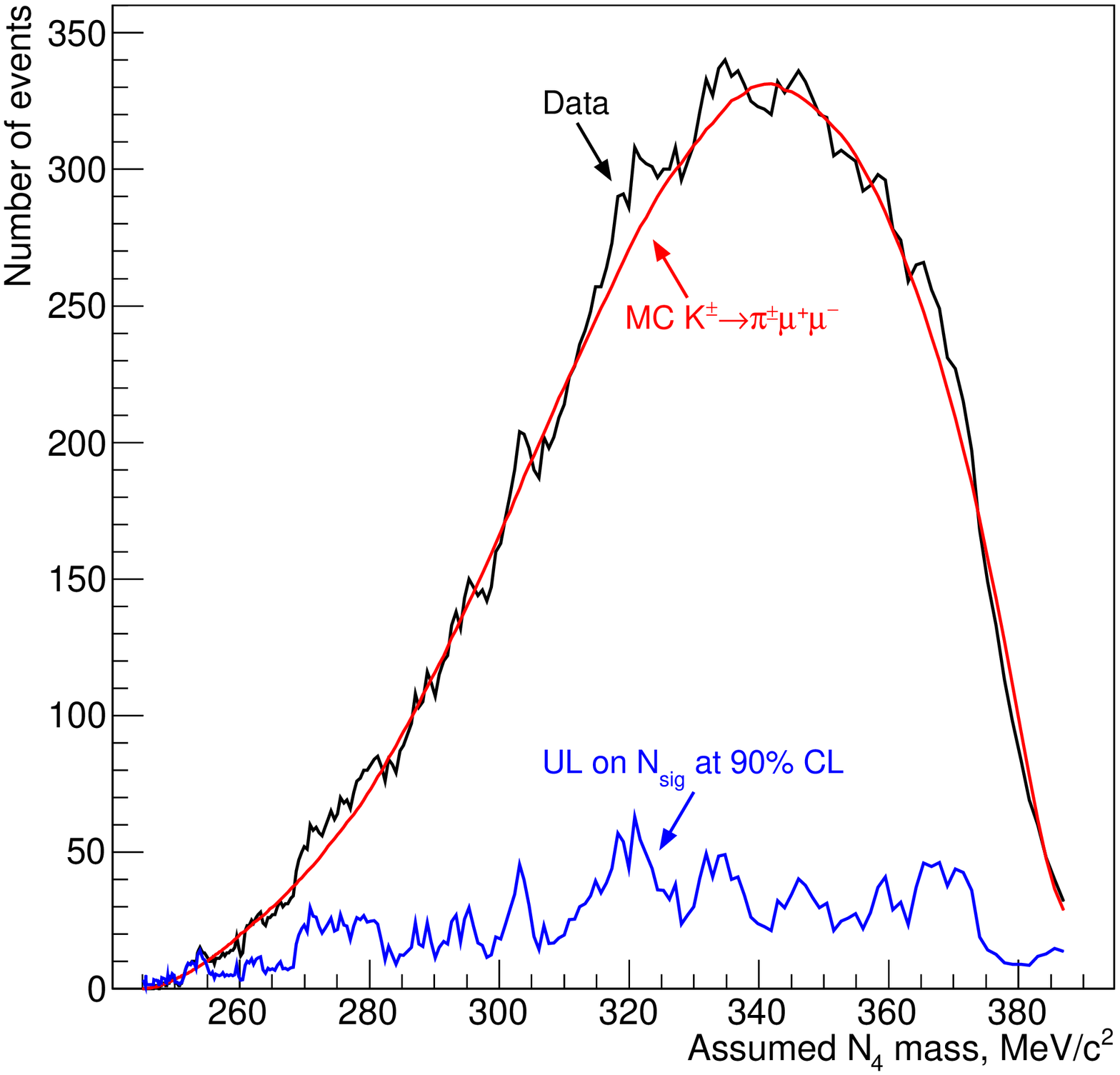}\\%
\hspace*{0.02\textwidth}\includegraphics[width=0.98\textwidth]{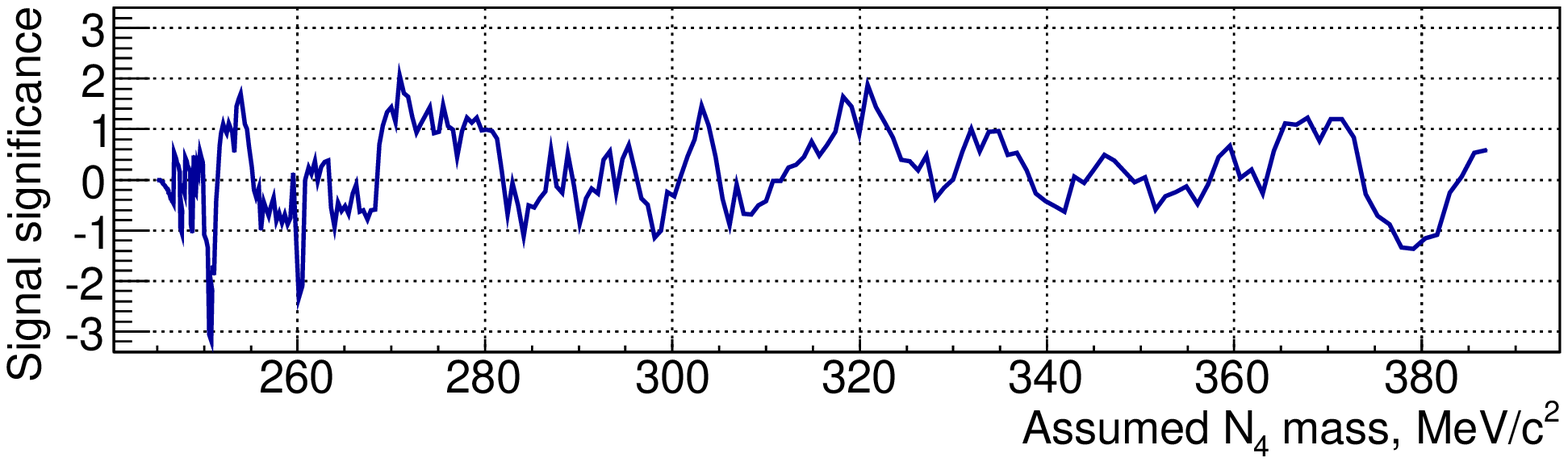}%
\end{minipage}
\put(-14,128){\Large\bf b}\\
\begin{minipage}{0.5\textwidth}
\includegraphics[width=\textwidth]{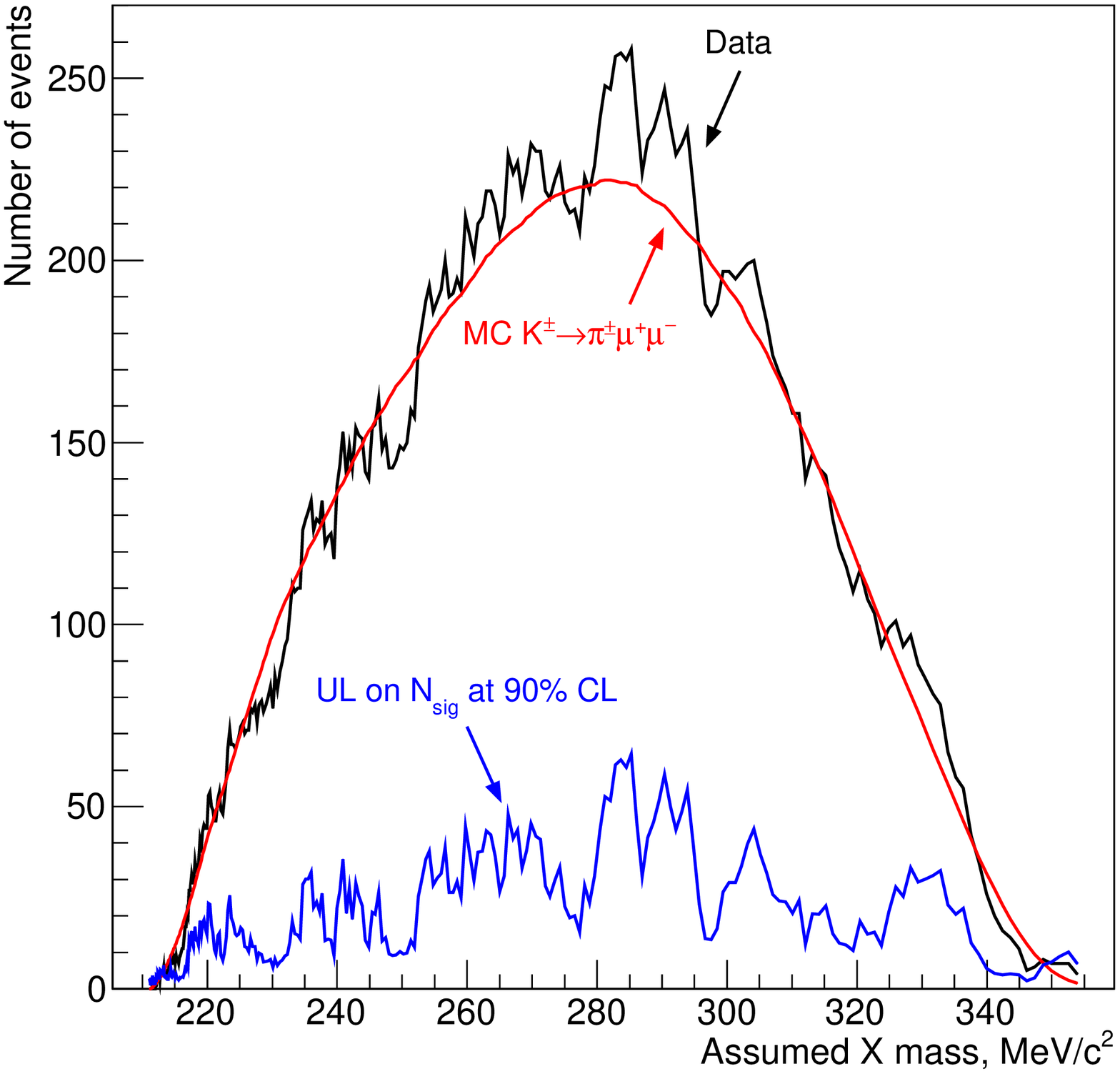}\\%
\hspace*{0.02\textwidth}\includegraphics[width=0.98\textwidth]{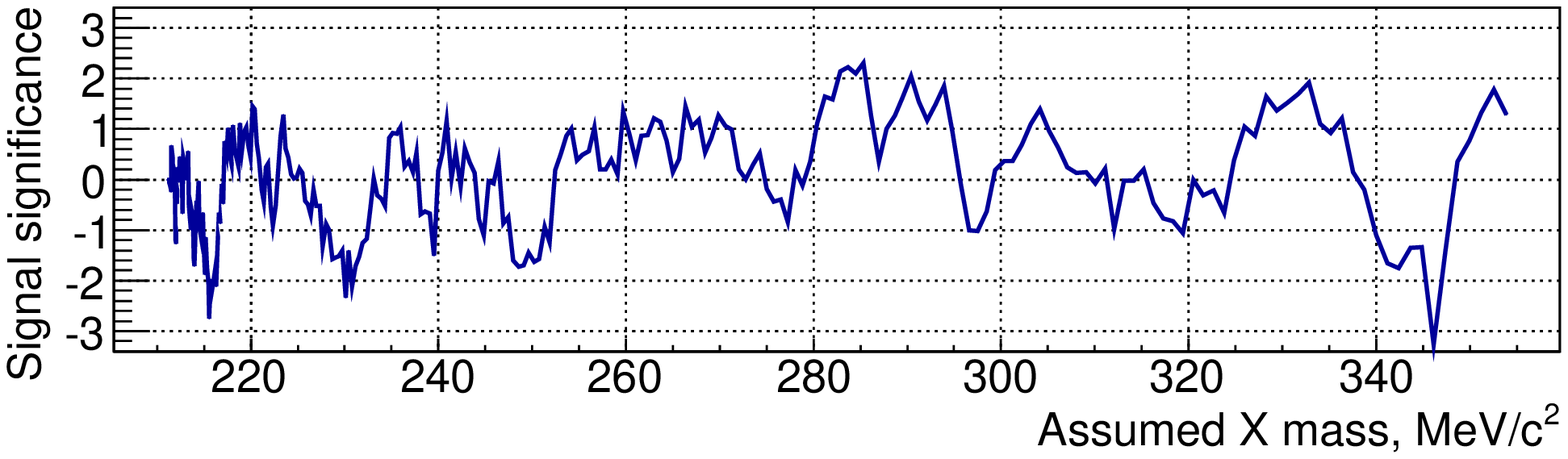}%
\end{minipage}
\put(-14,128){\Large\bf c}
\caption{Numbers of observed data events (black) and expected background events ($\kthreepic{\pm}$ in green, $\kpimm{\pm}$ in red) passing a) the $M_{\pi\mu}$ cut with the $\kpmmlnv$ selection; b) the $M_{\pi\mu}$ cut with the $\kpmmlnc$ selection; c) the $M_{\mu\mu}$ cut with the $\kpmmlnc$ selection.
The obtained upper limits at 90\% CL on the numbers of signal candidates (light blue) and the local significances of the signal (dark blue, in a separated figure) are also shown for each resonance mass value. All presented quantities are strongly correlated for neighbouring resonance masses as the mass step of the scan is about 8 times smaller than the signal window width.}\label{fig:nres_ul}  
\end{center}
\end{figure}

The statistical analysis of the obtained results in each mass window is performed by using a quasi-Newton minimisation algorithm to find numerically the 90\% confidence intervals for the case of a Poisson process in presence of unknown backgrounds, by applying an extension of the Rolke-Lopez method~\cite{ro01}.
For the generic case of $N$ considered backgrounds, the Rolke-Lopez computation performed requires $2N+1$ inputs for each mass hypothesis: the number~$N_{obs}$ of observed data events in the signal mass window; the number~$N_{bkg}^{\,i}$ of MC events for the considered background~$i$ observed in the signal mass window; the size~$\tau_i$, with respect to the data volume, of the MC sample used to evaluate~$N_{bkg}^{\,i}$ for the considered background~$i$.
The number of considered backgrounds for the $\kpmmlnv$ ($\kpmmlnc$) candidates is $N=4$ ($N=1$).
The values~$N_{obs}$, the normalised number of background events~$\tilde N_{bkg}^{\,i} \eqdef N_{bkg}^{\,i}/\tau_i$, 
and the upper limit~(UL) at 90\% confidence level~(CL) on the number~$N_{sig}$ of signal events obtained are shown for each mass hypothesis of the resonance searches in Fig.~\ref{fig:nres_ul}.

\boldmath
\section{Results}
\unboldmath
\label{sec:results}
\subsection{Upper Limit on $\mathcal{B}(\kpimmws)$}
The upper limit at 90\% CL on the number of $\kpimmws$ signal events in the $\kpmmlnv$ sample corresponding to the observation of one data event and a total number of expected background events $N_{bkg} = 1.163 \!\pm\! 0.867_{stat} \!\pm\! 0.021_{ext} \!\pm\! 0.116_{syst}$ is obtained applying the statistical analysis described in Section~\ref{sec:res_search} to the total number of events in the $\kpmmlnv$ sample: $\npmmlnv < 2.92$ at 90\%~CL. 
Using the values of the signal acceptance~$A(\kpmmlnv)=(20.62\pm0.01)\%$ estimated with MC simulations and the number~$N_K$ of kaon decays in the fiducial volume~(Section~\ref{sec:datasamples}),
the upper limit on the number of $\kpimmws$ signal events in the $\kpmmlnv$ sample leads to a constraint on the signal branching ratio~$\mathcal{B}(\kpimmws)$:
\begin{equation}
\label{eq:BR_kpimmws_experimental}
\mathcal{B}(\kpimmws) = \frac{\npmmlnv}{N_{3\pi}\!\cdot\! D}\cdot\frac{A(K_{3\pi})}{A(\kpmmlnv)}\cdot\mathcal{B}(K_{3\pi})< 8.6 \times 10^{-11} \quad \mbox{@ 90\% CL}.
\end{equation}
The total systematic uncertainty on the quoted upper limit is 1.5\%.
The largest source is the limi\-ted accuracy of the MC simulations (1.0\%), followed by $\mathcal{B}(\kpimm{\pm})$~(0.8\%),
$\mathcal{B}(\kthreepic{\pm})$~(0.73\%), $\mathcal{B}(\kmunumumu{\pm})$~(0.24\%) and $\mathcal{B}(\kmufour{\pm})$~(0.05\%).

\subsection{Results of the search for two-body resonances}
\label{subsubsec:n4mscan_ul}
For each of the three resonance searches performed, the local significance~$z$ of the signal has been evaluated for each mass hypothesis as
\begin{equation}
z = \frac{N_{obs}-N_{exp}}{\sqrt{\delta N^2_{obs}+\delta N^2_{exp}}},
\end{equation}
where $N_{obs}$ is the number of observed events, $N_{exp}$ is the number of expected background events, and $\delta {N_{obs}}$ ($\delta {N_{exp}}$) is the statistical uncertainty of $N_{obs}$ ($N_{exp}$).
The obtained results are shown in Fig.~\ref{fig:nres_ul}. No signal is observed, as the local significances never exceed 3 standard deviations.

\begin{figure}[p]
\begin{center}
\begin{minipage}{0.5\textwidth}
\includegraphics[width=\textwidth]{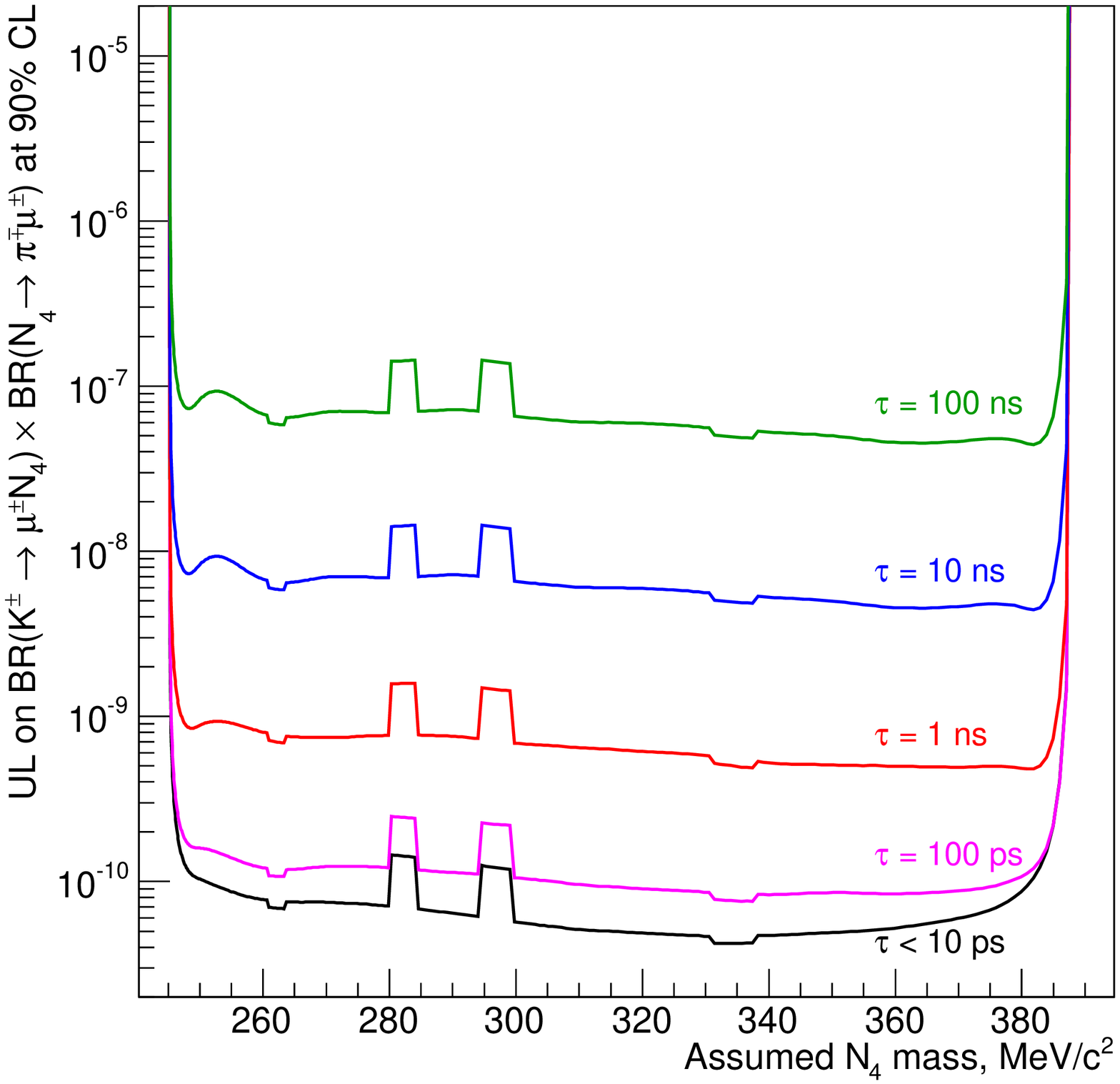}%
\end{minipage}
\put(-23,98){\Large\bf a}
\hfill
\begin{minipage}{0.5\textwidth}
\includegraphics[width=\textwidth]{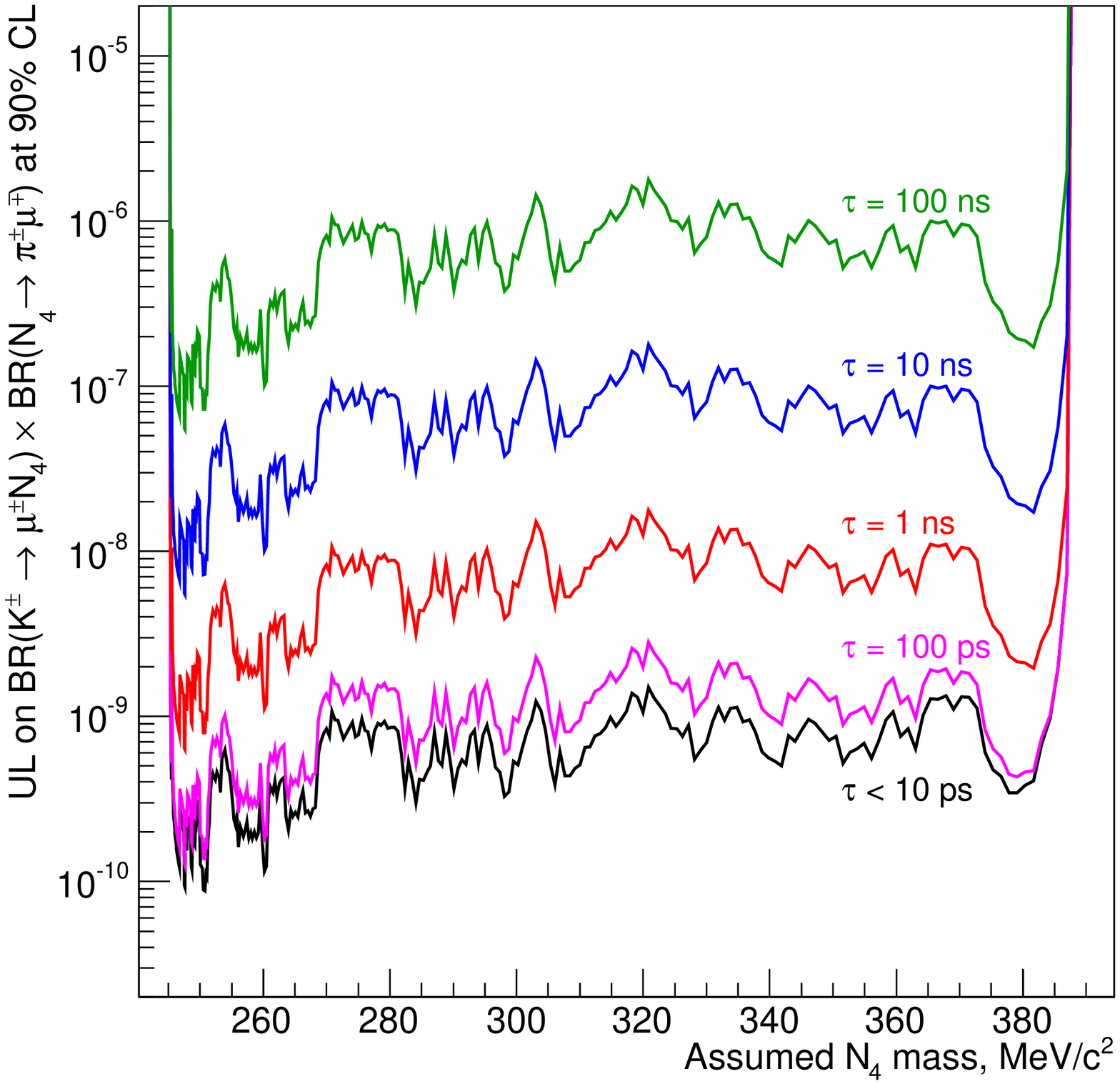}%
\end{minipage}
\put(-23,98){\Large\bf b}\\
\begin{minipage}{0.5\textwidth}
\includegraphics[width=\textwidth]{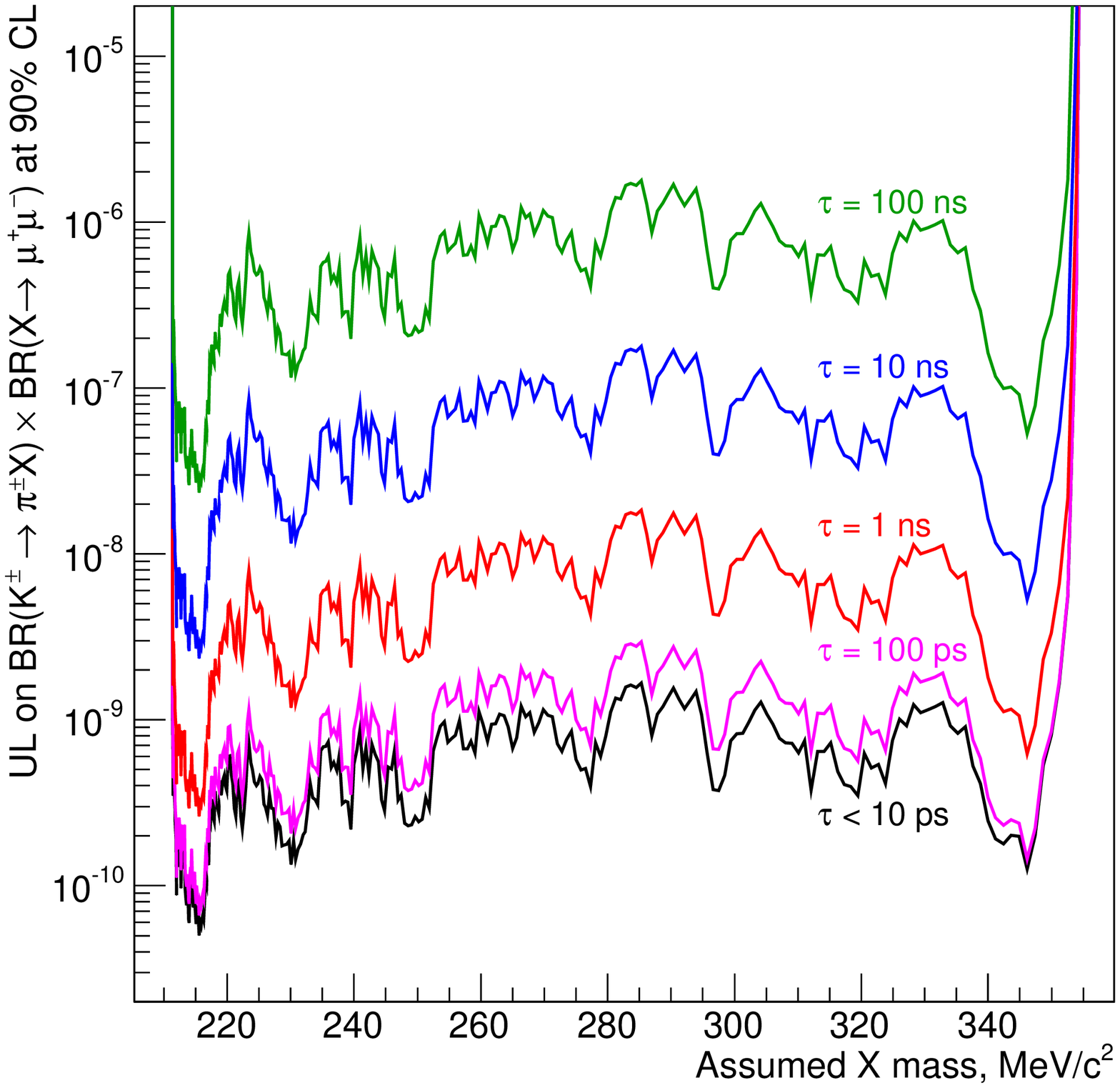}%
\end{minipage}
\put(-23,98){\Large\bf c}
\caption{Obtained upper limits at 90\% CL on the products of branching ratios as functions of the resonance mass and lifetime: a)~$\mathcal{B}(\kmutwoN{\pm})\mathcal{B}(\Npimuws)$; b)~$\mathcal{B}(\kmutwoN{\pm})\mathcal{B}(\Npimurs)$; c)~$\mathcal{B}(\kpichi{\pm})\mathcal{B}(\chimumu)$.
All presented quantities are strongly correlated for neighbouring resonance masses as the mass step of the scan is about 8 times smaller than the signal window width.}\label{fig:kpimmws_results_data}  
\end{center}
\end{figure}

In absence of a signal, upper limits on the product~$\mathcal{B}(K^{\pm}\to p_1 X)\mathcal{B}(X\to p_2 p_3)$ ($p_1p_2p_3 = \mu^{\pm}\pi^{\mp}\mu^{\pm},\mu^{\pm}\pi^{\pm}\mu^{\mp},\pi^{\pm}\mu^{+}\mu^{-}$) as a function of the resonance lifetime~$\tau$ are obtained for each mass hypothesis~$m_i$, by using the values of the acceptances~$A_{\pi\mu\mu}(m_i,\tau)$ (Fig.~\ref{fig:acc}) and the ULs on the number~$N^i_{sig}$ of signal events for such a mass hypothesis~(Fig.~\ref{fig:nres_ul}):
\begin{equation}
\label{eq:BR_res_massbin}
\left.\mathcal{B}(K^{\pm}\to p_1 X)\mathcal{B}(X\to p_2 p_3)\right|_{m_i,\tau} = \frac{N_{sig}^i}{N_{3\pi}\!\cdot\! D}\cdot\frac{A(K_{3\pi})}{A_{\pi\mu\mu}(m_i,\tau)}\cdot\mathcal{B}(K_{3\pi}).
\end{equation}
The obtained ULs on~$\mathcal{B}(K^{\pm}\to p_1 X)\mathcal{B}(X\to p_2 p_3)$ ($p_1p_2p_3 = \mu^{\pm}\pi^{\mp}\mu^{\pm},\mu^{\pm}\pi^{\pm}\mu^{\mp},\pi^{\pm}\mu^{+}\mu^{-}$) as a function of the resonance mass, for several values of the resonance lifetime, are shown in Fig.~\ref{fig:kpimmws_results_data}.

\section{Prospects for the new NA62 experiment}
The NA62 experiment exploits a major beam and detector upgrade to achieve the required physics specifications to perform the precise ($\sim 10\%$) measurement of the $\kpinn{+}$ branching ratio~\cite{ru16}.
A total of $\sim 1.5 \times 10^{13}$ $K^+$ decays is expected, with $\sim 3\times10^{12}$ $\pi^0$ decays from $K^{+}\to\pi^{+}\pi^0$. Studies of the prospects for searches for lepton-flavour (LF) or -number (LN) violating and other forbidden decays with NA62 are underway. 
The expected acceptance for rare $K^+$ ($\pi^0$) decays is $\sim 10\%$~$(1\%)$. Preliminary estimates of the single-event sensitivities (defined as the inverse of the number of accepted decays) give results at the level of $10^{-12}$ for $K^+$ decays to states such as $\pi^+\mu^{\pm}e^{\mp}$ (LFV), $\pi^-\mu^+e^+$ (LFNV), and $\pi^-e^+e^+$ or $\pi^-\mu^+\mu^+$ (LNV); and at the level of $10^{-11}$ for $\pi^0$ decays to~$\mu^{\pm} e^{\mp}$.

\section{Conclusions}
The searches for the LNV $\kpimmws$ decay and resonances in $\kpimmns{\pm}$ decays at the NA48/2 experiment, using the 2003--2004 data, are presented. 
No signals are observed. An upper limit of $8.6\times10^{-11}$ for the branching ratio of the LNV $\kpimmws$ decay has been established, which improves the best previous limit~\cite{ba11} by more than one order of magnitude.
Upper limits are set on the products of branching ratios~$\mathcal{B}(\kmutwoN{\pm})\mathcal{B}(\Npimuws)$ and $\mathcal{B}(\kpichi{\pm})\mathcal{B}(\chimumu)$ as functions of the resonance mass and lifetime. These limits are in the $10^{-10}-10^{-9}$ range for resonance lifetimes below 100~ps.

The forthcoming three-year long data taking period of the NA62 experiment will allow to search unexplored regions for heavy neutrinos, inflatons, and many forbidden kaon and $\pi^0$ decays, as the LNFV modes.

\end{document}